\documentclass[twocolumn,prc,aps,superscriptaddress,showpacs,floatfix]
{revtex4}

\def\be{\begin{eqnarray}}
\def\ee{\end{eqnarray}}
\def\bea{\begin{eqnarray}}
\def\eea{\end{eqnarray}}
\def\beas{\begin{eqnarray*}}
\def\eeas{\end{eqnarray*}}

\usepackage{graphics}
\usepackage{verbatim}
\usepackage{times}
\usepackage{epsfig}

\begin{document}

\bibliographystyle{apsrev}

\title{Nuclear Density Dependence of In-Medium Polarization}

\author{G. Ron}
\email{gron@phys.huji.ac.il}
\affiliation{Racah Institute of Physics, Hebrew University of Jerusalem, Givat
Ram 91904, Israel 
}

\author{W. Cosyn}
\email{wim.cosyn@ugent.be}
\affiliation{Department of Physics and Astronomy, Ghent University,
Proeftuinstraat 86, B-9000 Gent, Belgium}

\author{E. Piasetzky}
\email{eip@tauphy.tau.ac.il}
\affiliation{School of Physics and Astronomy, Tel Aviv University, Tel Aviv
69978, Israel }

\author{J. Ryckebusch}
\email{jan.ryckebusch@ugent.be}
\affiliation{Department of Physics and Astronomy, Ghent University,
Proeftuinstraat 86, B-9000 Gent, Belgium}

\author{J. Lichtenstadt}
\email{jech@tauphy.tau.ac.il}
\affiliation{School of Physics and Astronomy, Tel Aviv University, Tel Aviv
69978, Israel }

\date{\today}

\begin{abstract}
It is shown that polarization transfer measurements
$(\vec{e},e'\vec{p} \; )$ on a specific target nucleus can provide
constraints on the ratio of the in-medium electric to magnetic form
factor. Thereby one exploits the fact that proton knockout from
single-particle levels exhibit a specific sensitivity to the effective
nuclear density. It is shown that in $^{12}$C the effective nuclear
density for $s$-shell knockout is about twice as high as for $p$-shell
knockout. With current model predictions for the in-medium form
factors, one obtains measurable modifications of the order of 5\% in
the ratios of the double polarization observables between those
single-particle levels.
\end{abstract}


\maketitle 

Nuclei are well described as ensembles of protons and neutrons held
together by a strong mutual force.  The nucleons are complex entities
and the question of whether their internal structure is changed while
they are embedded in nuclei has been a long-standing question in
nuclear physics which remains unsettled~\cite{NuPECC}.

The polarization transfer measured in the $p(\vec{e},e'\vec{p} \; )$
reaction is a direct measure of the ratio of the proton elastic
electric to magnetic form factor (FF) ratio at some value of the
four-momentum transfer $Q ^ {2}$:
\begin{equation}
\frac{P_x}{P_z}=-\frac{2M_p}{(E+E')\tan(\theta/2)}
\frac{G_E^P \left(Q ^2 \right)}{G_M^P \left(Q ^2 \right)},
\end{equation}
where $P_x$ ($P_z$) is the transverse (longitudinal) polarization
transfer, $E$ ($E'$) is the incident (scattered) electron energy, $\theta$
is the electron scattering angle, and $M_p$ is the proton mass (see
~\cite{akhiezer74} for details).

When exclusive $(\vec{e},e'\vec{p} \; )$ measurements are performed
on a nuclear target, the polarization transfer observables are
sensitive to the modifications of the form factors of the embedded
nucleons, which we denote by
$\frac{G_E^*}{G_M^*}$~\cite{Cloet:2009tx,Strauch:2009wb,Paolone:2010qc}. The
double polarization ratios 
\begin{equation}
\left( \frac {\left( {P_x}/ {P_z} \right)_{A}} 
{\left( {P_x} / {P_z} \right)_{H}} \right), 
\end{equation}
taken between a knockout nucleon from a
nucleus $A$ and a free nucleon, are only moderately sensitive to many-body
effects like meson-exchange currents (MEC), isobar currents 
(IC), and final-state interactions
(FSI)~\cite{Laget:1994xx,Kelly:1998ti,ryckebusch99}. Small changes to
the measured observables in nuclei due to these many-body effects are
possible.  Distinguishing between the latter and the in-medium nucleon
structure modification is possible only using theoretical
calculations.  The challenge is to observe (or, exclude) deviations
which are outside the theoretical and experimental uncertainties that
can be used as evidence for changes in the bound nucleon form factor
compared to that of a free one.

The combination of high intensity, high polarization, continuous
electron beams, and high precision spectrometers with focal plane
polarimeters allows a measurement of the ratio of polarization
observables to a level of 1-2\%~\cite{Paolone:2010qc,Zhan:2011ji}.
With such measurements the theoretical uncertainties are the limiting
factor.

High-precision experiments and calculations, designed to look for
differences between the in-medium polarizations and the free values,
compared polarization observables measured in quasi-elastic scattering
off nuclear targets to these obtained for
hydrogen~\cite{Paolone:2010qc}. We discuss here the possibility to
identify in-medium effects and study their local nuclear density
dependence by comparing quasi-elastic proton removal from the
$s$-shell and $p$-shell in $^{12}$C. As we show below, in these cases
the local nuclear density is dramatically different.

Obtaining consistent results for medium modification if one compares
$s$-shell and $p$-shell knockout protons in the
$^{12}$C$(\vec{e},e'\vec{p} \; )^{11}$B reaction, or if one compares
the quasi-elastic scattering to that off a free proton, is a strong
support that can reduce the theoretical uncertainty of the magnitude
of the medium modifications. Moreover, one expects the medium
modification to depend on the local nuclear density and/or the bound
nucleon momentum/virtuality. Measurements that can map the effects as
a function of these two variables may reveal the nature of the medium
modifications.

The missing momentum $p_m$ corresponds to the initial momentum of the
struck nucleon in plane-wave kinematics.  In the deuteron, due to the
low nuclear density, the expected effect of medium modifications at
low missing momenta is too small to be detected
unambiguously~\cite{Eyl:1995fk,Milbrath:1997de,Barkhuff:1999xc,Hu:2006fy}. New
measurements for high missing momentum are still
unpublished~\cite{Israel}. Several polarization-transfer
proton-knockout experiments have been performed on $^4$He, both at the
MAMI facility~\cite{Dieterich:2000mu} and at Jefferson Lab
(JLab)~\cite{Strauch:2002wu,Paolone:2010qc}. The double-ratio of the
in-plane polarization components in $^{4}$He and a free proton,
\begin{equation} 
\left(\frac{(P_x/P_z)_{He}}{(P_x/P_z)_H}\right),
\label{eq:doubleration}
\end{equation}
which reflects the changes in the corresponding ratio of the electric
and magnetic form factors, does not agree with state-of-the-art
Distorted Wave Impulse Approximation (DWIA) calculations
~\cite{Paolone:2010qc} using free nucleon form factors, but can be
well described by including effects of medium modified form
factors~\cite{Lu:1997mu,PhysRevLett.91.212301,Smith:2004dn,
  Horikawa:2005dh,Lava:2005qp,Cloet:2005pp}. However, it has recently
been shown~\cite{PhysRevLett.94.072303} that including strong effects
from charge-exchange FSI can also explain the observed double-ratio of
Eq.~(\ref{eq:doubleration}).

The induced proton polarization in the $^{12}$C$(e,e'\vec{p} \; )$ reaction
has been reported by Woo~\cite{PhysRevLett.80.456} at quasi-elastic
kinematics and MAMI energies, covering a missing momentum range of
0--250~MeV/c. Polarization transfer measurements on $^{16}$O were
carried out at JLab~\cite{PhysRevC.62.057302}. Transverse and
longitudinal polarization components were measured in quasi-elastic
perpendicular kinematics at $Q^2$=0.8 (GeV/c)$^2$. The relatively
large uncertainties on both the measurements and the calculations did
not allow identification of deviations due to medium effects.

In this work we propose that the current state-of-the-art of
calculations and measurements allows observing possible medium
effects that are associated with local nuclear density. This can be
done by comparing quasi-elastic $s$-shell and $p$-shell removal of protons
from $^{12}$C rather than comparing quasi-elastic to elastic
scattering off Hydrogen.

We start by briefly presenting the Relativistic Multiple Scattering
Glauber Approximation (RMSGA). We then discuss the local nuclear
density difference between $s$- and $p$-shell protons in $^{12}$C and
present a few model calculations that estimate the magnitude of the
expected effect of medium modifications on double polarization
observables. We follow by proposing kinematics that can be accessed at
the MAMI/A1~\cite{Blomqvist1998263,Pospischil2002713} facility,
estimate the expected uncertainty, and compare it to the size of the
expected density dependent in-medium effect. Finally we discuss
the possible conclusions that can be drawn given the theoretical and
experimental limitations.

The Relativistic Multiple-Scattering Glauber Approximation
(RMSGA)~\cite{Ryckebusch:2003fc} is the theoretical framework used in
this work. It is a parameter-free model that was used to describe well
cross sections, nuclear transparencies and other observables in a large
variety of electron and hadron induced exclusive reactions in
kinematical conditions close to the case we discuss
here~\cite{Lava:2005qp,PhysRevC.83.054601,Cosyn:2010wk}. The RMSGA
provides an unfactorized approach to the $(e,e'p)$ reaction. In
contrast to factorized models which write the cross section in an
electron-proton part times a FSI-corrected nuclear-structure part, the
RMSGA computes the cross sections starting from the amplitudes. In the RMSGA 
the reaction amplitudes can be factorized in a part that describes the
wave function of the proton in the nuclear ground state, times an off-shell
current operator for the electron-proton scattering, times an
attenuation factor that accounts for the FSI of the emerging
proton. The eikonal Glauber FSI phase is a scalar in spin-space, hence the
FSI do not contain any spin effects.  The proton in the nuclear ground state is
described by a
single-particle bound state wave function obtained from the
Serot-Walecka model~\cite{Furnstahl:1996wv}.  To describe the
polarization observable in the polarized electron scattering off the
bound proton, the off-shell cross section CC2 was
used~\cite{DeForest:1983vc}. FSI were calculated using a relativistic
extension of the Glauber approximation. In the computation of the FSI,
the local nuclear density obtained from the independent-particle wave
function was corrected for the short-range correlations (SRC) assuming
Jastrow correlation function~\cite{Cosyn:2007er}.

\begin{figure}[ht]
\includegraphics[width=0.5\textwidth]{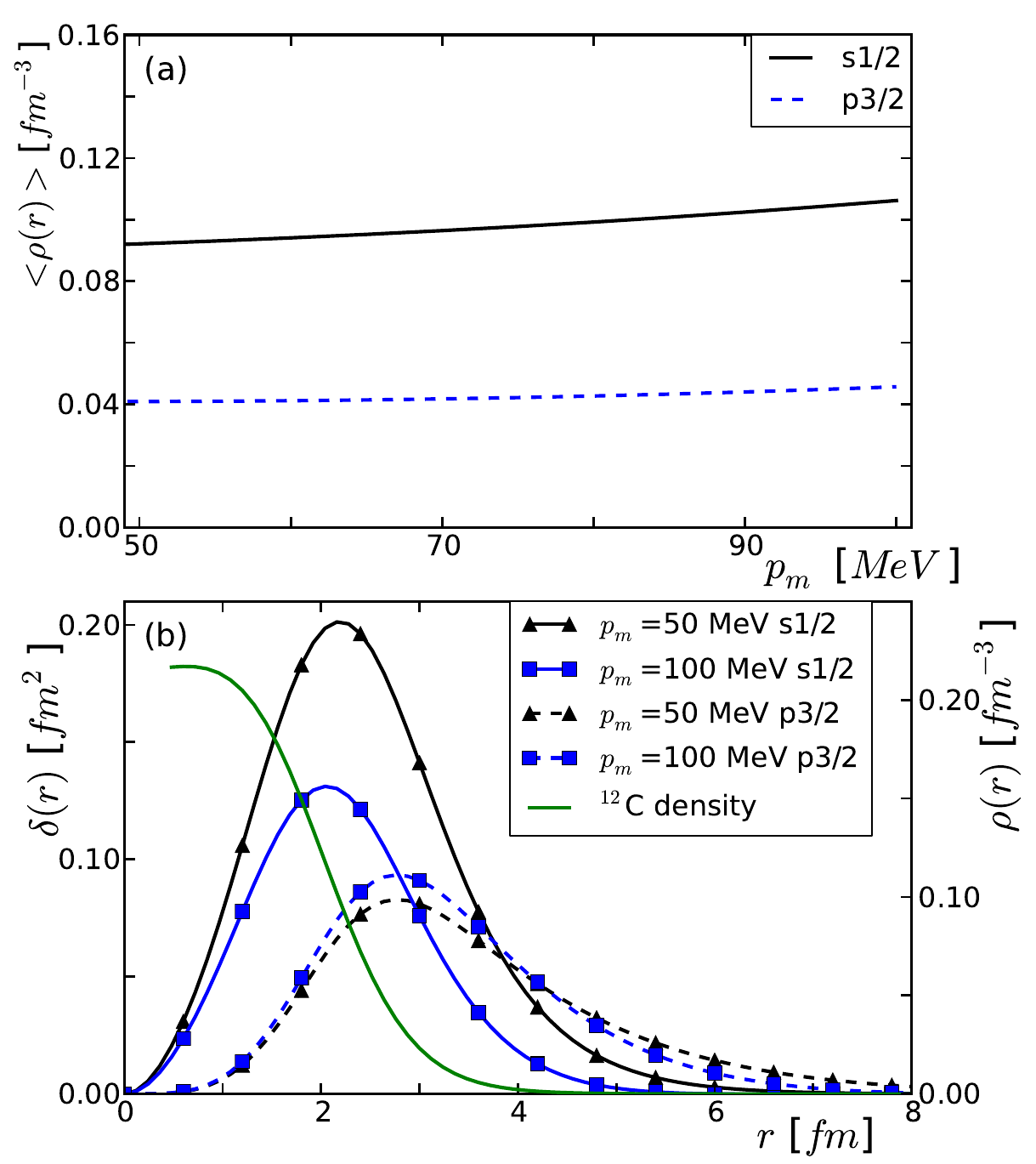}
\caption{\label{Fig:Density} (Color online) (a) Effective
  densities for protons removed from the $s$-shell and $p$-shell at
  $Q^2=0.4~(\text{GeV/c})^2$ as a function of missing momentum.
 (b) $\delta(r)$ for $s$-shell (full) and $p$-shell (dashed) removal
  for a missing momentum of 50 (black curves) and
  100 (blue curves) MeV.  The $^{12}$C density is also plotted (green curve) as a
  reference (scale on the right-side y-axis).
  }
\end{figure}

The effective density $\left< \rho (r) \right>$ for both the
$s$-shell and the $p$-shell proton in quasi-elastic proton knockout
from $^ {12}$C are shown in Fig.~\ref{Fig:Density} as a function of missing
momentum. These densities 
are obtained with mean-field single-particle wave functions.  We
observe the effective density probed in proton knockout from the
$s$-shell is about 0.1 fm$^{-3}$ and rises slightly with increasing
missing momentum $p_m$.  This is more than double the density probed
in knockout from the $p$-shell, which is around 0.04 fm$^{-3}$.  Also
shown in the figure is $\delta(r)$ which is the calculated contribution from
an infinitesimal density interval $[r,r+dr]$ to the cross section for a
quasifree $^{12}$C$(e,e'p)$ process and accounts for the effect of FSI
and SRC therein.  For a more detailed introduction on the quantity
$\delta(r)$ we refer to Refs.~\cite{Cosyn:2009bi,PhysRevC.83.054601}.
The FSI cause the largest contributions to the cross section to stem
from the peripheral regions of the proton densities.  These FSI
effects are strongest for the high-density regions of the nucleus and
thus affect the $s$-shell more than the $p$-shell.

\begin{figure}[ht]
\includegraphics[width=0.5\textwidth]{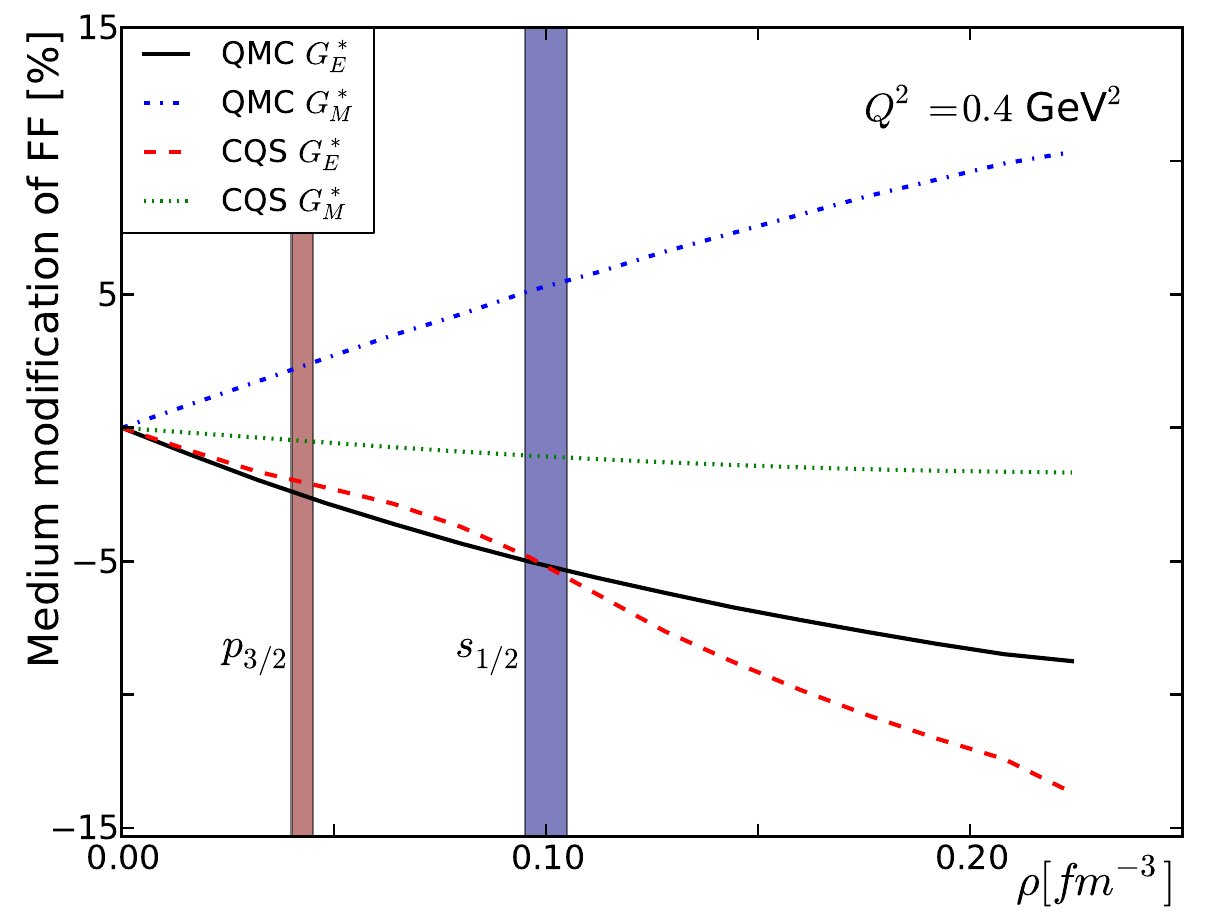}
\caption{\label{Fig:FFEM} (Color online) The nuclear density dependence of the
proton
  EM form factors from the QMC and CQS models as a function of nuclear density
at $Q^2 =$ 0.4
(GeV/c)$^2$.  The shaded bands show the effective nuclear densities for
the two proton shells probed  in
the $^{12}$C$(e,e'p)$ reaction at these kinematics.}
\end{figure}

To estimate the size of the in-medium modification we use two models
with density-dependent medium-modified elastic form factors for the
description of a bound proton. Fig.~\ref{Fig:FFEM} shows the nuclear
density dependence of the proton EM form factors at $Q^2 =$ 0.4
(GeV/c)$^2$ described by the two models.  In the Chiral Quark Soliton
(CQS) model~\cite{Christov:1995hr,Smith:2004dn} the sea quarks are
almost completely unaffected, whereas the valence quarks yield
significant modifications of the form factors in the nuclear
environment.  The model yields a decrease of the electric form factor
of about 5\% at nuclear saturation density ($\sim 0.16
~\text{fm}^{-3}$), while the modification of the magnetic form factor
is smaller, around 1-2\%.  In the Quark Meson Coupling (QMC)
model~\cite{Lu:1997mu,Lu:1998tn} the form factors are found to be
increasingly modified as the nuclear density increases. For example,
at saturation nuclear density, the nucleon electric form factor is,
reduced by approximately 7\%, similar to the CQS model. The magnetic
form factor increases by about the same amount, which is quite
different from the CQS value.

\begin{figure}[ht]
\begin{center}
                  \includegraphics[width=0.5\textwidth]{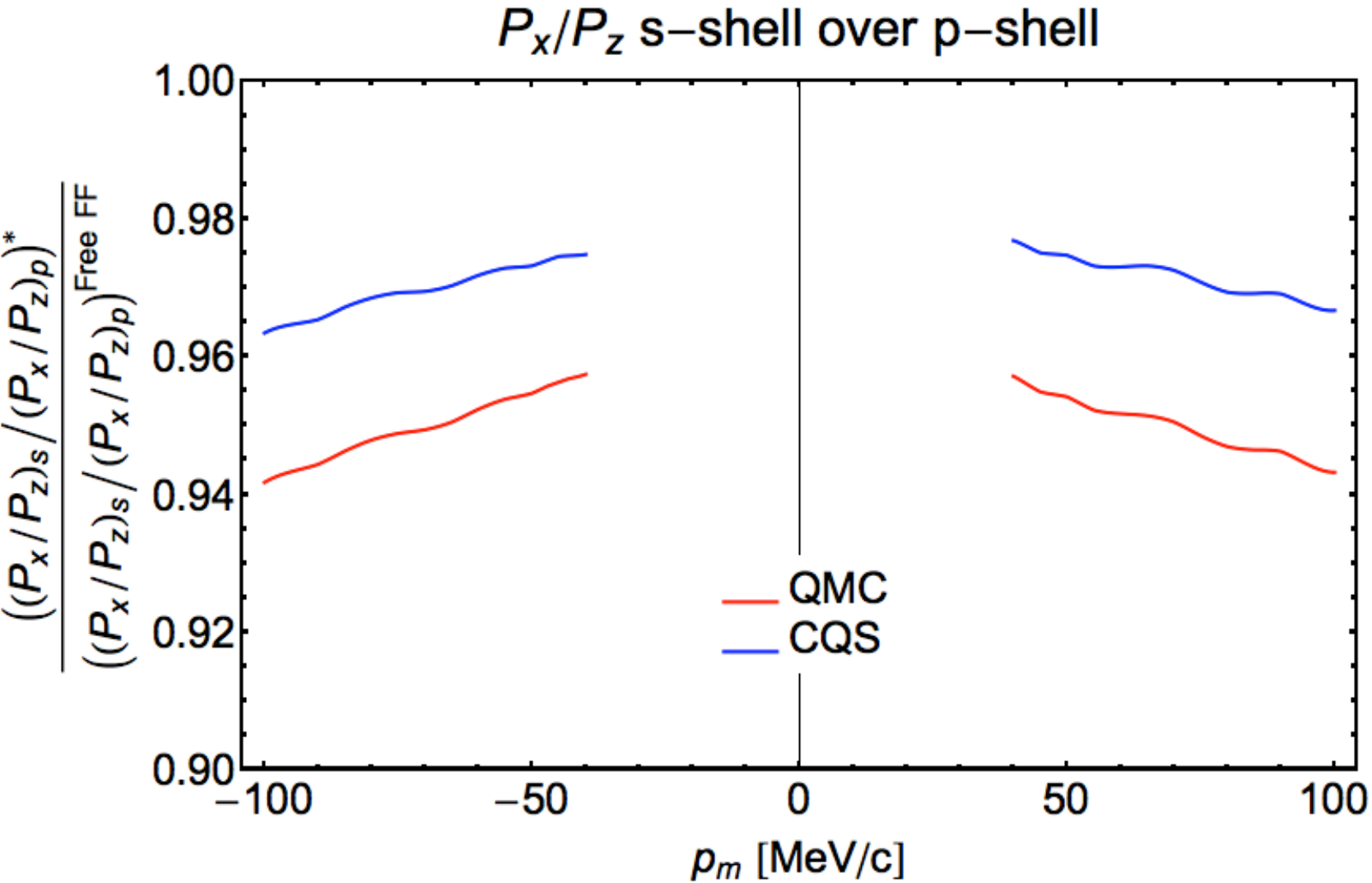}
\end{center}
\caption{(Color online) The ratio of the expected in-medium modification effect in
  the $s$- and $p$-shell removals. See text for details.}
\label{Fig:DoubRat}
\end{figure}

These QMC and CQS model calculations contained in Fig.~\ref{Fig:FFEM}
do not intend to yield precise predictions for the proposed
$^{12}$C$(e,e'\vec{p} \; )$
measurement, neither
to test/select the most appropriate model. These calculations point to
the possible size of the effect we expect to see from scattering off the
tightly bound $s$-shell proton relative to the less bound $p$-shell
proton. See Fig.~\ref{Fig:DoubRat} for an estimate of the difference
between the two shell removals for realistic measurement conditions
discussed below.

The suggested measurements can be performed using the MAMI/A1 beam
line and spectrometers~\cite{Blomqvist1998263,Pospischil2002713}. A 20
$\mu$A, 600 MeV, electron beam can be used to bombard a solid thin
carbon target. Two high resolution, small solid angle, spectrometers
will be used to detect the scattered electron and proton. The MAMI/A1
spectrometers have a scattering angle acceptance of approximately
$\pm$4 degrees, and a momentum acceptance of 20-25\%. The spectrometer
used to detect the proton is equipped with a focal plane polarimeter
(FPP) that is used to measure the polarization of the recoil
proton. The momentum resolution achievable by this setup allow
reconstructing the missing mass and clearly identifying the $s$- and $p$-
removal protons, which are separated by more than 2 MeV.

The proposed kinematics are, $Q^2$=0.4 (GeV/c)$^2$, a beam energy of
600~MeV, which gives a scattered electron energy of $E'$=384~MeV, and a
scattering angle of 82.4 (34.7) degree for the electron (proton). This
setup covers a missing momentum range of approximately
0$\pm$100~MeV/c.  At these kinematics the cross section is large enough
so that the data rate is limited by the Data Acquisition System. The
analyzing power of the FPP, and the spin precession angle of the
proton in the spectrometer magnetic field are such that within a
reasonable amount of beam time ($\sim$ 2 weeks) enough statistics can
be collected to ensure that the statistical uncertainties are smaller
than both the systematic and theoretical uncertainties. The expected
systematic uncertainties are dominated by the spin precession of the
proton in the magnetic field of the spectrometer, requiring an
accurate reconstruction of the proton trajectory in the magnetic
field, as well as knowledge of the field map. Comparison of the
measured polarization components with the well known results for a
free proton at the same Q$^2$ can be used to test the systematic
uncertainties. The false asymmetries are removed by using
straight-through runs, where the carbon analyzer is removed, resulting
in straight tracks throughout the polarimeter chambers. 
We estimate based on previous results~\cite{Paolone:2010qc,Zhan:2011ji,Dieterich:2000mu}, a conservative systematic
uncertainty of 2\% in the polarization ratio. Note, however, that this estimate is for the full acceptance of the spectrometer. 
The  comparison of the polarization ratios for s-shell and p-shell protons can be performed for individual parts of the focal 
plane and then combined. This procedure reduces the variation of the trajectories through the magnetic field, 
and its contribution to the systematic uncertainty.

Fig.~\ref{Fig:DoubRat} shows the predicted ratio of $s-$ and $p-$shell
removal calculations with in-medium modification to the modification
free ratio. The CQS and QMC models discussed above were used to
describe the in-medium case, the modification-free ratio was
calculated with free proton form factors (i.e., no medium
modification). All predictions use the RMSGA framework.  The ratio is
shown as a function of the $(e,e'p)$ missing momentum $p_m$ and
integrated over the acceptance of the MAMI/A1 spectrometers as listed
above. So Fig.~\ref{Fig:DoubRat} is our estimate of the result of the
proposed measurement.

In Fig.~\ref{Fig:DoubRat} super double ratios substantially different
from unity are an indication of in-medium modification.  As can be
deduced from Fig.~\ref{Fig:DoubRat} the expected effect is about 5\%.
With four $p_{m}$ bins (measured simultaneously) each measured with
1-2\% uncertainty, the deviation from unity can be determined with
very high certainty.

To summarize, we propose a measurement of the polarization transfer
components of the $s$-shell and $p$-shell knockout protons in the
$^{12}$C$(\vec{e},e'\vec{p} \; )^{11}$B reaction using the MAMI/A1
spectrometers. A free nucleon placed in the strong field of the
nucleus can have its structure modified. Proton recoil polarization,
measured in the quasi-elastic $^4$He$(\vec{e},e ^ {\prime} \vec{p} \;
)^3$H reaction, with unprecedented precision, were found to be
different from those of the $^1$H$(\vec{e},e ^ {\prime}\vec{p}\;)$
reaction, and studied as a function of proton virtuality.  Large
virtuality was also claimed to be the main cause of the EMC
effect~\cite{Ciofi:2007vx,Weinstein:2010rt}. Another possible source
of in-medium modification of the nucleon properties is the local
nuclear density probed in the reaction. Even at low missing momentum
one can expect off-shell behavior which most probably depends on the
nuclear density. In $^{12}$C, with high precision measurements, we can
identify proton removal from the $p$-shell and $s$-shell and study
possible changes between these two cases, where the local density is
more than a factor of two different.  In both the CQS and QMC model
our calculations show modification effects around 5\% for the double
ratio $\left(P_x/P_z \right)_{s} / \left(P_x/P_z \right) _{p}$ at the
proposed kinematics.

The authors wish to thank Harald Merkel and Micheal Distler for fruitful
discussions about the Mainz MAMI/AQ1 facility. This work is supported by the German Israeli Foundation,
Grant No.  2257-2141.14/2010, the Israel Science Foundation, Grant
No. 138/11, and the Research Foundation Flanders.  Computational
resources (Stevin Supercomputer Infrastructure) and services used in
this work were provided by Ghent University, the Hercules Foundation,
and the Flemish Government – department EWI.


\begin{thebibliography}{38}
\expandafter\ifx\csname natexlab\endcsname\relax\def\natexlab#1{#1}\fi
\expandafter\ifx\csname bibnamefont\endcsname\relax
  \def\bibnamefont#1{#1}\fi
\expandafter\ifx\csname bibfnamefont\endcsname\relax
  \def\bibfnamefont#1{#1}\fi
\expandafter\ifx\csname citenamefont\endcsname\relax
  \def\citenamefont#1{#1}\fi
\expandafter\ifx\csname url\endcsname\relax
  \def\url#1{\texttt{#1}}\fi
\expandafter\ifx\csname urlprefix\endcsname\relax\def\urlprefix{URL }\fi
\providecommand{\bibinfo}[2]{#2}
\providecommand{\eprint}[2][]{\url{#2}}

\bibitem[{\citenamefont{Bracco et~al.}(2010)\citenamefont{Bracco, Chomaz,
  Gaardh\o{}je, Heenen, Rosner, Widmann, and K\"{o}rner}}]{NuPECC}
\bibinfo{author}{\bibfnamefont{A.}~\bibnamefont{Bracco}},
  \bibinfo{author}{\bibfnamefont{P.}~\bibnamefont{Chomaz}},
  \bibinfo{author}{\bibfnamefont{J.}~\bibnamefont{Gaardh\o{}je}},
  \bibinfo{author}{\bibfnamefont{P.-H.} \bibnamefont{Heenen}},
  \bibinfo{author}{\bibfnamefont{G.}~\bibnamefont{Rosner}},
  \bibinfo{author}{\bibfnamefont{E.}~\bibnamefont{Widmann}}, \bibnamefont{and}
  \bibinfo{author}{\bibfnamefont{G.-E.} \bibnamefont{K\"{o}rner}},
  \emph{\bibinfo{title}{{NuPECC Long Range Plan 2010: Perspectives of Nuclear
  Physics in Europe}}} (\bibinfo{year}{2010}),
  \urlprefix\url{http://www.nupecc.org/}.

\bibitem[{\citenamefont{Akhiezer and Rekalo}(1974)}]{akhiezer74}
\bibinfo{author}{\bibfnamefont{A.~I.} \bibnamefont{Akhiezer}} \bibnamefont{and}
  \bibinfo{author}{\bibfnamefont{M.~P.} \bibnamefont{Rekalo}},
  \bibinfo{journal}{Sov. J. Part. Nucl.} \textbf{\bibinfo{volume}{4}},
  \bibinfo{pages}{277} (\bibinfo{year}{1974}).

\bibitem[{\citenamefont{Cloet et~al.}(2009)\citenamefont{Cloet, Miller,
  Piasetzky, and Ron}}]{Cloet:2009tx}
\bibinfo{author}{\bibfnamefont{I.}~\bibnamefont{Cloet}},
  \bibinfo{author}{\bibfnamefont{G.~A.} \bibnamefont{Miller}},
  \bibinfo{author}{\bibfnamefont{E.}~\bibnamefont{Piasetzky}},
  \bibnamefont{and} \bibinfo{author}{\bibfnamefont{G.}~\bibnamefont{Ron}},
  \bibinfo{journal}{Phys. Rev. Lett.} \textbf{\bibinfo{volume}{103}},
  \bibinfo{pages}{082301} (\bibinfo{year}{2009}).

\bibitem[{\citenamefont{Strauch et~al.}(2009)\citenamefont{Strauch, Malace, and
  Paolone}}]{Strauch:2009wb}
\bibinfo{author}{\bibfnamefont{S.}~\bibnamefont{Strauch}},
  \bibinfo{author}{\bibfnamefont{S.}~\bibnamefont{Malace}}, \bibnamefont{and}
  \bibinfo{author}{\bibfnamefont{M.}~\bibnamefont{Paolone}}
  (\bibinfo{year}{2009}).

\bibitem[{\citenamefont{Paolone et~al.}(2010)\citenamefont{Paolone, Malace,
  Strauch, Albayrak, Arrington, Berman, Brash, Briscoe, Camsonne, Chen
  et~al.}}]{Paolone:2010qc}
\bibinfo{author}{\bibfnamefont{M.}~\bibnamefont{Paolone}},
  \bibinfo{author}{\bibfnamefont{S.~P.} \bibnamefont{Malace}},
  \bibinfo{author}{\bibfnamefont{S.}~\bibnamefont{Strauch}},
  \bibinfo{author}{\bibfnamefont{I.}~\bibnamefont{Albayrak}},
  \bibinfo{author}{\bibfnamefont{J.}~\bibnamefont{Arrington}},
  \bibinfo{author}{\bibfnamefont{B.~L.} \bibnamefont{Berman}},
  \bibinfo{author}{\bibfnamefont{E.~J.} \bibnamefont{Brash}},
  \bibinfo{author}{\bibfnamefont{B.}~\bibnamefont{Briscoe}},
  \bibinfo{author}{\bibfnamefont{A.}~\bibnamefont{Camsonne}},
  \bibinfo{author}{\bibfnamefont{J.-P.} \bibnamefont{Chen}},
  \bibnamefont{et~al.} (\bibinfo{collaboration}{E03-104 Collaboration}),
  \bibinfo{journal}{Phys. Rev. Lett.} \textbf{\bibinfo{volume}{105}},
  \bibinfo{pages}{072001} (\bibinfo{year}{2010}).

\bibitem[{\citenamefont{Laget}(1994)}]{Laget:1994xx}
\bibinfo{author}{\bibfnamefont{J.-M.} \bibnamefont{Laget}},
  \bibinfo{journal}{Nucl. Phys.} \textbf{\bibinfo{volume}{A579}},
  \bibinfo{pages}{333} (\bibinfo{year}{1994}).

\bibitem[{\citenamefont{Kelly}(1999)}]{Kelly:1998ti}
\bibinfo{author}{\bibfnamefont{J.~J.} \bibnamefont{Kelly}},
  \bibinfo{journal}{Phys. Rev.} \textbf{\bibinfo{volume}{C59}},
  \bibinfo{pages}{3256} (\bibinfo{year}{1999}).

\bibitem[{\citenamefont{Ryckebusch et~al.}(1999)\citenamefont{Ryckebusch,
  Debruyne, {Van Nespen}, and Janssen}}]{ryckebusch99}
\bibinfo{author}{\bibfnamefont{J.}~\bibnamefont{Ryckebusch}},
  \bibinfo{author}{\bibfnamefont{D.}~\bibnamefont{Debruyne}},
  \bibinfo{author}{\bibfnamefont{W.}~\bibnamefont{{Van Nespen}}},
  \bibnamefont{and} \bibinfo{author}{\bibfnamefont{S.}~\bibnamefont{Janssen}},
  \bibinfo{journal}{Phys. Rev.} \textbf{\bibinfo{volume}{C60}},
  \bibinfo{pages}{034604} (\bibinfo{year}{1999}).

\bibitem[{\citenamefont{Zhan et~al.}(2011)\citenamefont{Zhan, Allada,
  Armstrong, Arrington, Bertozzi, Boeglin, Chen, Chirapatpimol, Choi, Chudakov
  et~al.}}]{Zhan:2011ji}
\bibinfo{author}{\bibfnamefont{X.}~\bibnamefont{Zhan}},
  \bibinfo{author}{\bibfnamefont{K.}~\bibnamefont{Allada}},
  \bibinfo{author}{\bibfnamefont{D.}~\bibnamefont{Armstrong}},
  \bibinfo{author}{\bibfnamefont{J.}~\bibnamefont{Arrington}},
  \bibinfo{author}{\bibfnamefont{W.}~\bibnamefont{Bertozzi}},
  \bibinfo{author}{\bibfnamefont{W.}~\bibnamefont{Boeglin}},
  \bibinfo{author}{\bibfnamefont{J.-P.} \bibnamefont{Chen}},
  \bibinfo{author}{\bibfnamefont{K.}~\bibnamefont{Chirapatpimol}},
  \bibinfo{author}{\bibfnamefont{S.}~\bibnamefont{Choi}},
  \bibinfo{author}{\bibfnamefont{E.}~\bibnamefont{Chudakov}},
  \bibnamefont{et~al.}, \bibinfo{journal}{Phys. Lett.}
  \textbf{\bibinfo{volume}{B705}}, \bibinfo{pages}{59} (\bibinfo{year}{2011}).

\bibitem[{\citenamefont{Eyl et~al.}(1995)\citenamefont{Eyl, Frey, Andresen,
  Annand, Aulenbacher, Becker, Blume-Werry, Dombo, Drescher, Fischer
  et~al.}}]{Eyl:1995fk}
\bibinfo{author}{\bibfnamefont{D.}~\bibnamefont{Eyl}},
  \bibinfo{author}{\bibfnamefont{A.}~\bibnamefont{Frey}},
  \bibinfo{author}{\bibfnamefont{H.}~\bibnamefont{Andresen}},
  \bibinfo{author}{\bibfnamefont{J.}~\bibnamefont{Annand}},
  \bibinfo{author}{\bibfnamefont{K.}~\bibnamefont{Aulenbacher}},
  \bibinfo{author}{\bibfnamefont{J.}~\bibnamefont{Becker}},
  \bibinfo{author}{\bibfnamefont{J.}~\bibnamefont{Blume-Werry}},
  \bibinfo{author}{\bibfnamefont{T.}~\bibnamefont{Dombo}},
  \bibinfo{author}{\bibfnamefont{P.}~\bibnamefont{Drescher}},
  \bibinfo{author}{\bibfnamefont{H.}~\bibnamefont{Fischer}},
  \bibnamefont{et~al.}, \bibinfo{journal}{Z. Phys.}
  \textbf{\bibinfo{volume}{A352}}, \bibinfo{pages}{211} (\bibinfo{year}{1995}).

\bibitem[{\citenamefont{Milbrath et~al.}(1998)\citenamefont{Milbrath, McIntyre,
  Armstrong, Barkhuff, Bertozzi, Chen, Dale, Dodson, Dow, Epstein
  et~al.}}]{Milbrath:1997de}
\bibinfo{author}{\bibfnamefont{B.~D.} \bibnamefont{Milbrath}},
  \bibinfo{author}{\bibfnamefont{J.~I.} \bibnamefont{McIntyre}},
  \bibinfo{author}{\bibfnamefont{C.~S.} \bibnamefont{Armstrong}},
  \bibinfo{author}{\bibfnamefont{D.~H.} \bibnamefont{Barkhuff}},
  \bibinfo{author}{\bibfnamefont{W.}~\bibnamefont{Bertozzi}},
  \bibinfo{author}{\bibfnamefont{J.~P.} \bibnamefont{Chen}},
  \bibinfo{author}{\bibfnamefont{D.}~\bibnamefont{Dale}},
  \bibinfo{author}{\bibfnamefont{G.}~\bibnamefont{Dodson}},
  \bibinfo{author}{\bibfnamefont{K.~A.} \bibnamefont{Dow}},
  \bibinfo{author}{\bibfnamefont{M.~B.} \bibnamefont{Epstein}},
  \bibnamefont{et~al.} (\bibinfo{collaboration}{Bates FPP Collaboration}),
  \bibinfo{journal}{Phys. Rev. Lett.} \textbf{\bibinfo{volume}{80}},
  \bibinfo{pages}{452} (\bibinfo{year}{1998}).

\bibitem[{\citenamefont{Barkhuff et~al.}(1999)\citenamefont{Barkhuff,
  Armstrong, Bertozzi, Chen, Dale, Dodson, Dow, Epstein, Farkhondeh, Finn
  et~al.}}]{Barkhuff:1999xc}
\bibinfo{author}{\bibfnamefont{D.}~\bibnamefont{Barkhuff}},
  \bibinfo{author}{\bibfnamefont{C.}~\bibnamefont{Armstrong}},
  \bibinfo{author}{\bibfnamefont{W.}~\bibnamefont{Bertozzi}},
  \bibinfo{author}{\bibfnamefont{J.}~\bibnamefont{Chen}},
  \bibinfo{author}{\bibfnamefont{D.}~\bibnamefont{Dale}},
  \bibinfo{author}{\bibfnamefont{G.}~\bibnamefont{Dodson}},
  \bibinfo{author}{\bibfnamefont{K.}~\bibnamefont{Dow}},
  \bibinfo{author}{\bibfnamefont{M.}~\bibnamefont{Epstein}},
  \bibinfo{author}{\bibfnamefont{M.}~\bibnamefont{Farkhondeh}},
  \bibinfo{author}{\bibfnamefont{J.}~\bibnamefont{Finn}}, \bibnamefont{et~al.},
  \bibinfo{journal}{Phys. Lett.} \textbf{\bibinfo{volume}{B470}},
  \bibinfo{pages}{39} (\bibinfo{year}{1999}).

\bibitem[{\citenamefont{Hu et~al.}(2006)\citenamefont{Hu, Jones, Ulmer,
  Arenh\"ovel, Baker, Bertozzi, Brash, Calarco, Chen, Chudakov
  et~al.}}]{Hu:2006fy}
\bibinfo{author}{\bibfnamefont{B.}~\bibnamefont{Hu}},
  \bibinfo{author}{\bibfnamefont{M.~K.} \bibnamefont{Jones}},
  \bibinfo{author}{\bibfnamefont{P.~E.} \bibnamefont{Ulmer}},
  \bibinfo{author}{\bibfnamefont{H.}~\bibnamefont{Arenh\"ovel}},
  \bibinfo{author}{\bibfnamefont{O.~K.} \bibnamefont{Baker}},
  \bibinfo{author}{\bibfnamefont{W.}~\bibnamefont{Bertozzi}},
  \bibinfo{author}{\bibfnamefont{E.~J.} \bibnamefont{Brash}},
  \bibinfo{author}{\bibfnamefont{J.}~\bibnamefont{Calarco}},
  \bibinfo{author}{\bibfnamefont{J.-P.} \bibnamefont{Chen}},
  \bibinfo{author}{\bibfnamefont{E.}~\bibnamefont{Chudakov}},
  \bibnamefont{et~al.}, \bibinfo{journal}{Phys. Rev. C}
  \textbf{\bibinfo{volume}{73}}, \bibinfo{pages}{064004}
  (\bibinfo{year}{2006}).

\bibitem[{\citenamefont{Yaron et~al.}(2013)}]{Israel}
\bibinfo{author}{\bibfnamefont{I.}~\bibnamefont{Yaron}} \bibnamefont{et~al.},
  \bibinfo{journal}{To be submitted}  (\bibinfo{year}{2013}).

\bibitem[{\citenamefont{Dieterich et~al.}(2001)\citenamefont{Dieterich,
  Bartsch, Baumann, Bermuth, Bohinc, Böhm, Bosnar, Derber, Ding, Distler
  et~al.}}]{Dieterich:2000mu}
\bibinfo{author}{\bibfnamefont{S.}~\bibnamefont{Dieterich}},
  \bibinfo{author}{\bibfnamefont{P.}~\bibnamefont{Bartsch}},
  \bibinfo{author}{\bibfnamefont{D.}~\bibnamefont{Baumann}},
  \bibinfo{author}{\bibfnamefont{J.}~\bibnamefont{Bermuth}},
  \bibinfo{author}{\bibfnamefont{K.}~\bibnamefont{Bohinc}},
  \bibinfo{author}{\bibfnamefont{R.}~\bibnamefont{Böhm}},
  \bibinfo{author}{\bibfnamefont{D.}~\bibnamefont{Bosnar}},
  \bibinfo{author}{\bibfnamefont{S.}~\bibnamefont{Derber}},
  \bibinfo{author}{\bibfnamefont{M.}~\bibnamefont{Ding}},
  \bibinfo{author}{\bibfnamefont{M.}~\bibnamefont{Distler}},
  \bibnamefont{et~al.}, \bibinfo{journal}{Phys. Lett.}
  \textbf{\bibinfo{volume}{B500}}, \bibinfo{pages}{47} (\bibinfo{year}{2001}).
  
\bibitem[{\citenamefont{Strauch et~al.}(2003)\citenamefont{Strauch, Dieterich,
  Aniol, Annand, Baker, Bertozzi, Boswell, Brash, Chai, Chen
  et~al.}}]{Strauch:2002wu}
\bibinfo{author}{\bibfnamefont{S.}~\bibnamefont{Strauch}},
  \bibinfo{author}{\bibfnamefont{S.}~\bibnamefont{Dieterich}},
  \bibinfo{author}{\bibfnamefont{K.~A.} \bibnamefont{Aniol}},
  \bibinfo{author}{\bibfnamefont{J.~R.~M.} \bibnamefont{Annand}},
  \bibinfo{author}{\bibfnamefont{O.~K.} \bibnamefont{Baker}},
  \bibinfo{author}{\bibfnamefont{W.}~\bibnamefont{Bertozzi}},
  \bibinfo{author}{\bibfnamefont{M.}~\bibnamefont{Boswell}},
  \bibinfo{author}{\bibfnamefont{E.~J.} \bibnamefont{Brash}},
  \bibinfo{author}{\bibfnamefont{Z.}~\bibnamefont{Chai}},
  \bibinfo{author}{\bibfnamefont{J.-P.} \bibnamefont{Chen}},
  \bibnamefont{et~al.}, \bibinfo{journal}{Phys. Rev. Lett.}
  \textbf{\bibinfo{volume}{91}}, \bibinfo{pages}{052301}
  (\bibinfo{year}{2003}).

\bibitem[{\citenamefont{Lu et~al.}(1998)\citenamefont{Lu, Thomas, Tsushima,
  Williams, and Saito}}]{Lu:1997mu}
\bibinfo{author}{\bibfnamefont{D.-H.} \bibnamefont{Lu}},
  \bibinfo{author}{\bibfnamefont{A.~W.} \bibnamefont{Thomas}},
  \bibinfo{author}{\bibfnamefont{K.}~\bibnamefont{Tsushima}},
  \bibinfo{author}{\bibfnamefont{A.~G.} \bibnamefont{Williams}},
  \bibnamefont{and} \bibinfo{author}{\bibfnamefont{K.}~\bibnamefont{Saito}},
  \bibinfo{journal}{Phys. Lett.} \textbf{\bibinfo{volume}{B417}},
  \bibinfo{pages}{217} (\bibinfo{year}{1998}).

\bibitem[{\citenamefont{Smith and Miller}(2003)}]{PhysRevLett.91.212301}
\bibinfo{author}{\bibfnamefont{J.~R.} \bibnamefont{Smith}} \bibnamefont{and}
  \bibinfo{author}{\bibfnamefont{G.~A.} \bibnamefont{Miller}},
  \bibinfo{journal}{Phys. Rev. Lett.} \textbf{\bibinfo{volume}{91}},
  \bibinfo{pages}{212301} (\bibinfo{year}{2003}).

\bibitem[{\citenamefont{Smith and Miller}(2004)}]{Smith:2004dn}
\bibinfo{author}{\bibfnamefont{J.~R.} \bibnamefont{Smith}} \bibnamefont{and}
  \bibinfo{author}{\bibfnamefont{G.~A.} \bibnamefont{Miller}},
  \bibinfo{journal}{Phys. Rev.} \textbf{\bibinfo{volume}{C70}},
  \bibinfo{pages}{065205} (\bibinfo{year}{2004}).

\bibitem[{\citenamefont{Horikawa and Bentz}(2005)}]{Horikawa:2005dh}
\bibinfo{author}{\bibfnamefont{T.}~\bibnamefont{Horikawa}} \bibnamefont{and}
  \bibinfo{author}{\bibfnamefont{W.}~\bibnamefont{Bentz}},
  \bibinfo{journal}{Nucl. Phys.} \textbf{\bibinfo{volume}{A762}},
  \bibinfo{pages}{102} (\bibinfo{year}{2005}).

\bibitem[{\citenamefont{Lava et~al.}(2005)\citenamefont{Lava, Ryckebusch, and
  Van~Overmeire}}]{Lava:2005qp}
\bibinfo{author}{\bibfnamefont{P.}~\bibnamefont{Lava}},
  \bibinfo{author}{\bibfnamefont{J.}~\bibnamefont{Ryckebusch}},
  \bibnamefont{and}
  \bibinfo{author}{\bibfnamefont{B.}~\bibnamefont{Van~Overmeire}},
  \bibinfo{journal}{Prog. Part. Nucl. Phys.} \textbf{\bibinfo{volume}{55}},
  \bibinfo{pages}{437} (\bibinfo{year}{2005}).

\bibitem[{\citenamefont{Cloet et~al.}(2005)\citenamefont{Cloet, Bentz, and
  Thomas}}]{Cloet:2005pp}
\bibinfo{author}{\bibfnamefont{I.}~\bibnamefont{Cloet}},
  \bibinfo{author}{\bibfnamefont{W.}~\bibnamefont{Bentz}}, \bibnamefont{and}
  \bibinfo{author}{\bibfnamefont{A.~W.} \bibnamefont{Thomas}},
  \bibinfo{journal}{Phys. Lett.} \textbf{\bibinfo{volume}{B621}},
  \bibinfo{pages}{246} (\bibinfo{year}{2005}).

\bibitem[{\citenamefont{Schiavilla et~al.}(2005)\citenamefont{Schiavilla,
  Benhar, Kievsky, Marcucci, and Viviani}}]{PhysRevLett.94.072303}
\bibinfo{author}{\bibfnamefont{R.}~\bibnamefont{Schiavilla}},
  \bibinfo{author}{\bibfnamefont{O.}~\bibnamefont{Benhar}},
  \bibinfo{author}{\bibfnamefont{A.}~\bibnamefont{Kievsky}},
  \bibinfo{author}{\bibfnamefont{L.~E.} \bibnamefont{Marcucci}},
  \bibnamefont{and} \bibinfo{author}{\bibfnamefont{M.}~\bibnamefont{Viviani}},
  \bibinfo{journal}{Phys. Rev. Lett.} \textbf{\bibinfo{volume}{94}},
  \bibinfo{pages}{072303} (\bibinfo{year}{2005}).

\bibitem[{\citenamefont{Woo et~al.}(1998)\citenamefont{Woo, Barkhuff, Bertozzi,
  Chen, Dale, Dodson, Dow, Epstein, Farkhondeh, Finn
  et~al.}}]{PhysRevLett.80.456}
\bibinfo{author}{\bibfnamefont{R.~J.} \bibnamefont{Woo}},
  \bibinfo{author}{\bibfnamefont{D.~H.} \bibnamefont{Barkhuff}},
  \bibinfo{author}{\bibfnamefont{W.}~\bibnamefont{Bertozzi}},
  \bibinfo{author}{\bibfnamefont{J.~P.} \bibnamefont{Chen}},
  \bibinfo{author}{\bibfnamefont{D.}~\bibnamefont{Dale}},
  \bibinfo{author}{\bibfnamefont{G.}~\bibnamefont{Dodson}},
  \bibinfo{author}{\bibfnamefont{K.~A.} \bibnamefont{Dow}},
  \bibinfo{author}{\bibfnamefont{M.~B.} \bibnamefont{Epstein}},
  \bibinfo{author}{\bibfnamefont{M.}~\bibnamefont{Farkhondeh}},
  \bibinfo{author}{\bibfnamefont{J.~M.} \bibnamefont{Finn}},
  \bibnamefont{et~al.} (\bibinfo{collaboration}{Bates FPP Collaboration}),
  \bibinfo{journal}{Phys. Rev. Lett.} \textbf{\bibinfo{volume}{80}},
  \bibinfo{pages}{456} (\bibinfo{year}{1998}).

\bibitem[{\citenamefont{Malov et~al.}(2000)\citenamefont{Malov, Wijesooriya,
  Baker, Bimbot, Brash, Chang, Finn, Fissum, Gao, Gilman
  et~al.}}]{PhysRevC.62.057302}
\bibinfo{author}{\bibfnamefont{S.}~\bibnamefont{Malov}},
  \bibinfo{author}{\bibfnamefont{K.}~\bibnamefont{Wijesooriya}},
  \bibinfo{author}{\bibfnamefont{F.~T.} \bibnamefont{Baker}},
  \bibinfo{author}{\bibfnamefont{L.}~\bibnamefont{Bimbot}},
  \bibinfo{author}{\bibfnamefont{E.~J.} \bibnamefont{Brash}},
  \bibinfo{author}{\bibfnamefont{C.~C.} \bibnamefont{Chang}},
  \bibinfo{author}{\bibfnamefont{J.~M.} \bibnamefont{Finn}},
  \bibinfo{author}{\bibfnamefont{K.~G.} \bibnamefont{Fissum}},
  \bibinfo{author}{\bibfnamefont{J.}~\bibnamefont{Gao}},
  \bibinfo{author}{\bibfnamefont{R.}~\bibnamefont{Gilman}},
  \bibnamefont{et~al.}, \bibinfo{journal}{Phys. Rev. C}
  \textbf{\bibinfo{volume}{62}}, \bibinfo{pages}{057302}
  (\bibinfo{year}{2000}).

\bibitem[{\citenamefont{Blomqvist et~al.}(1998)\citenamefont{Blomqvist,
  Boeglin, Bšhm, Distler, Edelhoff, Friedrich, Geiges, Jennewein, Kahrau, Korn
  et~al.}}]{Blomqvist1998263}
\bibinfo{author}{\bibfnamefont{K.}~\bibnamefont{Blomqvist}},
  \bibinfo{author}{\bibfnamefont{W.}~\bibnamefont{Boeglin}},
  \bibinfo{author}{\bibfnamefont{R.}~\bibnamefont{Bšhm}},
  \bibinfo{author}{\bibfnamefont{M.}~\bibnamefont{Distler}},
  \bibinfo{author}{\bibfnamefont{R.}~\bibnamefont{Edelhoff}},
  \bibinfo{author}{\bibfnamefont{J.}~\bibnamefont{Friedrich}},
  \bibinfo{author}{\bibfnamefont{R.}~\bibnamefont{Geiges}},
  \bibinfo{author}{\bibfnamefont{P.}~\bibnamefont{Jennewein}},
  \bibinfo{author}{\bibfnamefont{M.}~\bibnamefont{Kahrau}},
  \bibinfo{author}{\bibfnamefont{M.}~\bibnamefont{Korn}}, \bibnamefont{et~al.},
  \bibinfo{journal}{Nuclear Instruments and Methods in Physics Research Section
  A: Accelerators, Spectrometers, Detectors and Associated Equipment}
  \textbf{\bibinfo{volume}{403}}, \bibinfo{pages}{263 } (\bibinfo{year}{1998}).
  
\bibitem[{\citenamefont{Pospischil et~al.}(2002)\citenamefont{Pospischil,
  Bartsch, Baumann, Bšhm, Bohinc, Clawiter, Ding, Derber, Distler, Elsner
  et~al.}}]{Pospischil2002713}
\bibinfo{author}{\bibfnamefont{T.}~\bibnamefont{Pospischil}},
  \bibinfo{author}{\bibfnamefont{P.}~\bibnamefont{Bartsch}},
  \bibinfo{author}{\bibfnamefont{D.}~\bibnamefont{Baumann}},
  \bibinfo{author}{\bibfnamefont{R.}~\bibnamefont{Bšhm}},
  \bibinfo{author}{\bibfnamefont{K.}~\bibnamefont{Bohinc}},
  \bibinfo{author}{\bibfnamefont{N.}~\bibnamefont{Clawiter}},
  \bibinfo{author}{\bibfnamefont{M.}~\bibnamefont{Ding}},
  \bibinfo{author}{\bibfnamefont{S.}~\bibnamefont{Derber}},
  \bibinfo{author}{\bibfnamefont{M.}~\bibnamefont{Distler}},
  \bibinfo{author}{\bibfnamefont{D.}~\bibnamefont{Elsner}},
  \bibnamefont{et~al.}, \bibinfo{journal}{Nuclear Instruments and Methods in
  Physics Research Section A: Accelerators, Spectrometers, Detectors and
  Associated Equipment} \textbf{\bibinfo{volume}{483}}, \bibinfo{pages}{713 }
  (\bibinfo{year}{2002}).

\bibitem[{\citenamefont{Ryckebusch et~al.}(2003)\citenamefont{Ryckebusch,
  Debruyne, Lava, Janssen, Van~Overmeire, and
  Van~Cauteren}}]{Ryckebusch:2003fc}
\bibinfo{author}{\bibfnamefont{J.}~\bibnamefont{Ryckebusch}},
  \bibinfo{author}{\bibfnamefont{D.}~\bibnamefont{Debruyne}},
  \bibinfo{author}{\bibfnamefont{P.}~\bibnamefont{Lava}},
  \bibinfo{author}{\bibfnamefont{S.}~\bibnamefont{Janssen}},
  \bibinfo{author}{\bibfnamefont{B.}~\bibnamefont{Van~Overmeire}},
  \bibnamefont{and}
  \bibinfo{author}{\bibfnamefont{T.}~\bibnamefont{Van~Cauteren}},
  \bibinfo{journal}{Nucl. Phys.} \textbf{\bibinfo{volume}{A728}},
  \bibinfo{pages}{226} (\bibinfo{year}{2003}).

\bibitem[{\citenamefont{Ryckebusch et~al.}(2011)\citenamefont{Ryckebusch,
  Cosyn, and Vanhalst}}]{PhysRevC.83.054601}
\bibinfo{author}{\bibfnamefont{J.}~\bibnamefont{Ryckebusch}},
  \bibinfo{author}{\bibfnamefont{W.}~\bibnamefont{Cosyn}}, \bibnamefont{and}
  \bibinfo{author}{\bibfnamefont{M.}~\bibnamefont{Vanhalst}},
  \bibinfo{journal}{Phys. Rev. C} \textbf{\bibinfo{volume}{83}},
  \bibinfo{pages}{054601} (\bibinfo{year}{2011}).

\bibitem[{\citenamefont{Cosyn and Ryckebusch}(2011)}]{Cosyn:2010wk}
\bibinfo{author}{\bibfnamefont{W.}~\bibnamefont{Cosyn}} \bibnamefont{and}
  \bibinfo{author}{\bibfnamefont{J.}~\bibnamefont{Ryckebusch}},
  \bibinfo{journal}{Few Body Syst.} \textbf{\bibinfo{volume}{49}},
  \bibinfo{pages}{77} (\bibinfo{year}{2011}).

\bibitem[{\citenamefont{Furnstahl et~al.}(1997)\citenamefont{Furnstahl, Serot,
  and Tang}}]{Furnstahl:1996wv}
\bibinfo{author}{\bibfnamefont{R.}~\bibnamefont{Furnstahl}},
  \bibinfo{author}{\bibfnamefont{B.~D.} \bibnamefont{Serot}}, \bibnamefont{and}
  \bibinfo{author}{\bibfnamefont{H.-B.} \bibnamefont{Tang}},
  \bibinfo{journal}{Nucl. Phys.} \textbf{\bibinfo{volume}{A615}},
  \bibinfo{pages}{441} (\bibinfo{year}{1997}).

\bibitem[{\citenamefont{De~Forest}(1983)}]{DeForest:1983vc}
\bibinfo{author}{\bibfnamefont{T.}~\bibnamefont{De~Forest}},
  \bibinfo{journal}{Nucl. Phys.} \textbf{\bibinfo{volume}{A392}},
  \bibinfo{pages}{232} (\bibinfo{year}{1983}).

\bibitem[{\citenamefont{Cosyn et~al.}(2008)\citenamefont{Cosyn, Martinez, and
  Ryckebusch}}]{Cosyn:2007er}
\bibinfo{author}{\bibfnamefont{W.}~\bibnamefont{Cosyn}},
  \bibinfo{author}{\bibfnamefont{M.~C.} \bibnamefont{Martinez}},
  \bibnamefont{and}
  \bibinfo{author}{\bibfnamefont{J.}~\bibnamefont{Ryckebusch}},
  \bibinfo{journal}{Phys. Rev. C} \textbf{\bibinfo{volume}{77}},
  \bibinfo{pages}{034602} (\bibinfo{year}{2008}).

\bibitem[{\citenamefont{Cosyn and Ryckebusch}(2009)}]{Cosyn:2009bi}
\bibinfo{author}{\bibfnamefont{W.}~\bibnamefont{Cosyn}} \bibnamefont{and}
  \bibinfo{author}{\bibfnamefont{J.}~\bibnamefont{Ryckebusch}},
  \bibinfo{journal}{Phys. Rev.} \textbf{\bibinfo{volume}{C80}},
  \bibinfo{pages}{011602} (\bibinfo{year}{2009}).

\bibitem[{\citenamefont{Christov et~al.}(1995)\citenamefont{Christov, Gorski,
  Goeke, and Pobylitsa}}]{Christov:1995hr}
\bibinfo{author}{\bibfnamefont{C.}~\bibnamefont{Christov}},
  \bibinfo{author}{\bibfnamefont{A.}~\bibnamefont{Gorski}},
  \bibinfo{author}{\bibfnamefont{K.}~\bibnamefont{Goeke}}, \bibnamefont{and}
  \bibinfo{author}{\bibfnamefont{P.}~\bibnamefont{Pobylitsa}},
  \bibinfo{journal}{Nucl. Phys.} \textbf{\bibinfo{volume}{A592}},
  \bibinfo{pages}{513} (\bibinfo{year}{1995}).

\bibitem[{\citenamefont{Lu et~al.}(1999)\citenamefont{Lu, Tsushima, Thomas,
  Williams, and Saito}}]{Lu:1998tn}
\bibinfo{author}{\bibfnamefont{D.-H.} \bibnamefont{Lu}},
  \bibinfo{author}{\bibfnamefont{K.}~\bibnamefont{Tsushima}},
  \bibinfo{author}{\bibfnamefont{A.~W.} \bibnamefont{Thomas}},
  \bibinfo{author}{\bibfnamefont{A.~G.} \bibnamefont{Williams}},
  \bibnamefont{and} \bibinfo{author}{\bibfnamefont{K.}~\bibnamefont{Saito}},
  \bibinfo{journal}{Phys. Rev.} \textbf{\bibinfo{volume}{C60}},
  \bibinfo{pages}{068201} (\bibinfo{year}{1999}).

\bibitem[{\citenamefont{Ciofi~degli Atti et~al.}(2007)\citenamefont{Ciofi~degli
  Atti, Frankfurt, Kaptari, and Strikman}}]{Ciofi:2007vx}
\bibinfo{author}{\bibfnamefont{C.}~\bibnamefont{Ciofi~degli Atti}},
  \bibinfo{author}{\bibfnamefont{L.}~\bibnamefont{Frankfurt}},
  \bibinfo{author}{\bibfnamefont{L.}~\bibnamefont{Kaptari}}, \bibnamefont{and}
  \bibinfo{author}{\bibfnamefont{M.}~\bibnamefont{Strikman}},
  \bibinfo{journal}{Phys. Rev.} \textbf{\bibinfo{volume}{C76}},
  \bibinfo{pages}{055206} (\bibinfo{year}{2007}).

\bibitem[{\citenamefont{Weinstein et~al.}(2011)\citenamefont{Weinstein,
  Piasetzky, Higinbotham, Gomez, Hen, and Shneor}}]{Weinstein:2010rt}
\bibinfo{author}{\bibfnamefont{L.~B.} \bibnamefont{Weinstein}},
  \bibinfo{author}{\bibfnamefont{E.}~\bibnamefont{Piasetzky}},
  \bibinfo{author}{\bibfnamefont{D.~W.} \bibnamefont{Higinbotham}},
  \bibinfo{author}{\bibfnamefont{J.}~\bibnamefont{Gomez}},
  \bibinfo{author}{\bibfnamefont{O.}~\bibnamefont{Hen}}, \bibnamefont{and}
  \bibinfo{author}{\bibfnamefont{R.}~\bibnamefont{Shneor}},
  \bibinfo{journal}{Phys. Rev. Lett.} \textbf{\bibinfo{volume}{106}},
  \bibinfo{pages}{052301} (\bibinfo{year}{2011}).

\end{thebibliography}
\end{document}